# Super-reflection and Cloaking Based on Zero Index Metamaterial

Jiaming Hao, Wei Yan, and Min Qiu*

*Photonics and Microwave Engineering, Royal Institute of Technology (KTH),*
*Electrum 229, 164 40, Kista, Sweden*

**Abstract**

A zero index metamaterial (ZIM) can be utilized to block wave (super-reflection) or conceal objects completely (cloaking). The "super-reflection" device is realized by a ZIM with a perfect electric (magnetic) conductor inclusion of arbitrary shape and size for a transverse electric (magnetic) incident wave. In contrast, a ZIM with a perfect magnetic (electric) conductor inclusion for a transverse electric (magnetic) incident wave can be used to conceal objects of arbitrary shape. The underlying physics here is determined by the intrinsic properties of the ZIM.

Metamaterials with unusually tailored values of electric permittivity and magnetic permeability have attracted tremendous attention recently [1, 2]. Especially, combining with the transformation optics method, a lot of intriguing wave-functional devices have been proposed, for example invisibility cloaks [3, 4], electromagnetic field concentrators [5], field rotators [6], cylindrical superlenses [7] and superscatterer [8]. On the other hand, zero index metamaterials (ZIMs) whose permittivity and permeability are simultaneously or individually equal to zero, as a specific type of metamaterials, have also become a vigorous topic of scientific research [9-15]. Enoch *et al.* [9] presented that such metamaterial can be used to enhance the directivity of the emission of an embedded source; Li *et al.* [10] showed that zero volume averaged refractive index materials display photonic band gaps with properties different from that of a Bragg gap; Silveirinha and Engheta [12, 13] proposed that electromagnetic waves can be "squeezed" and tunneled through very narrow channels filled with $\varepsilon$ -near-zero materials, which has been demonstrated experimentally at microwave frequency regime [14, 15]. In this letter, we propose that a ZIM can be used to block wave with an arbitrary small inclusion (super-reflection) or conceal objects completely (cloaking) under certain conditions. The underlying physics here are determined by the intrinsic properties of ZIMs, i.e. in steady state the electromagnetic fields in the ZIM material are static without phase variation in space.

Consider a two-dimensional waveguide structure, which consists of four distinct regions, as illustrated in Fig. 1. The region (0) and region (3) are filled with air separated by a ZIM (region (1)) with effective electromagnetic parameters $\varepsilon_1, \mu_1$. An inclusion (region (2)) with relative electromagnetic parameters $\varepsilon_2, \mu_2$ is embedded in the host medium ZIM. The walls of waveguide are made of perfect magnetic conductor (PMC), and the waveguide supports the fundamental transverse electric (TE) wave mode. (Results are similar for TM waves if one switches to perfect electric conductor (PEC).) Suppose that a TE wave with $\vec{E}_{in} = E_{0i}\hat{z}e^{i(k_0x-\omega t)}$ impinging upon the surface of ZIM. For convenience, we omit the time variation item in the rest of this letter and ignore the effect of losses. Thus we know that the electric and magnetic field in the region (0) can be written as

$$\vec{E}^{(0)} = E_{0i}\hat{z}e^{ik_0x} + RE_{0i}\hat{z}e^{-ik_0x} \tag{1}$$

$$\vec{H}^{(0)} = \frac{1}{i\omega\mu_0}\nabla\times\vec{E}^{(0)} = \frac{E_{0i}}{i\omega\mu_0}\hat{y}(-ik_0e^{ik_0x} + ik_0Re^{-ik_0x}) \tag{2}$$

while the electromagnetic field in the region (3) must have the form

$$\vec{E}^{(3)} = TE_{0i}\hat{z}e^{ik_0(x-a)} \tag{3}$$

$$\vec{H}^{(3)} = \frac{1}{i\omega\mu_0}\nabla\times\vec{E}^{(3)} = \frac{1}{i\omega\mu_0}(-ik_0)TE_{0i}\hat{y}e^{ik_0(x-a)} \tag{4}$$

where $R$ and $T$ are the reflection and transmission coefficients for the electric field, respectively. We note that in the region (2) the electric field should satisfy the following differential equation

$$\nabla\times\left(\frac{1}{\mu_2}\nabla\times\vec{E}^{(2)}\right)+k_0^2\varepsilon_2\vec{E}^{(2)} = 0\,, \tag{5}$$

and subject to the boundary condition $E^{(2)}\big|_{\partial A_2} = E^{(1)}$ at the boundary of the region (2), $\partial A_2$. The magnetic field in the interior of the region (2) is

$$\vec{H}^{(2)} = \frac{1}{i\omega\mu_0\mu_2}\nabla\times\vec{E}^{(2)} \tag{6}$$

In the region (1), the magnetic field is also given by $\vec{H}^{(1)} = \frac{1}{i\omega\mu_0\mu_1}\nabla\times\vec{E}^{(1)}$. But, in the limit of $\mu_1 \approx 0$, in order to keep $\vec{H}^{(1)}$ as finite value, the electric field $\vec{E}^{(1)}$ must be constant in the ZIM. Appling the boundary condition, the tangential components of electromagnetic fields should be continuous at the interfaces $x = 0$ and $x = a$,

$$E^{(1)} = (1+R)E_{0i}\big|_{x=0} \tag{7}$$

$$H^{(1)} = \frac{E_{0i}}{i\omega\mu_0}(-ik_0+ik_0R) = \frac{1}{\eta_0}(R-1)E_{0i}\bigg|_{x=0} \tag{8}$$

$$E^{(3)} = TE_{0i} = E^{(1)}\big|_{x=a} \tag{9}$$

$$H^{(3)} = -\frac{1}{\eta_0}TE_{0i}\bigg|_{x=a} \tag{10}$$

where $\eta_0 = \sqrt{\mu_0/\varepsilon_0}$ is free space impedance. Due to $\vec{E}^{(1)}$ being constant and using Eqs (7) and (9) one has

$$T = 1+R \tag{11}$$

Next, we use Ampére-Maxwell law of magnetic fields [16] to solve the unknown scattering parameters,

$$\oint \vec{H}\cdot d\vec{l} = I_f + \int \frac{\partial\vec{D}}{\partial t}\cdot d\vec{S} \tag{12}$$

where $I_f$ denotes the total free current, which vanishes in our system. For ZIM medium (Region 1), the left-hand integral in Eq. (12) can be written as

$$\oint_{\partial A_1} \vec{H}\cdot d\vec{l} = \frac{E_{0i}}{\eta_0}\left(2dR-(R+1)\frac{\eta_0}{E^{(1)}}\oint_{\partial A2}\vec{H}^{(1)}\cdot d\vec{l}\right), \tag{13}$$

where $\partial A_1$ and $\partial A_2$ represent the boundary of the region (1) and (2), respectively. The right-hand integral is given by

$$\int_{A_1}\frac{\partial\vec{D}}{\partial t}\cdot d\vec{S} = -i\omega\varepsilon_0\varepsilon_1(1+R)E_{0i}A_1, \tag{14}$$

where $A_1$ denote the area of the region (1). Using Eqs (13) and (14), it is easy to find that the reflection and transmission coefficient can be written as following [9, 10]

$$R = \frac{-ik_0\varepsilon_1 A_1 + \frac{\eta_0}{E^{(1)}}\oint_{\partial A2}\vec{H}^{(1)}\cdot d\vec{l}}{2d + ik_0\varepsilon_1 A_1 - \frac{\eta_0}{E^{(1)}}\oint_{\partial A2}\vec{H}^{(1)}\cdot d\vec{l}}, \tag{15}$$

$$T = \frac{2d}{2d + ik_0\varepsilon_1 A_1 - \frac{\eta_0}{E^{(1)}}\oint_{\partial A2}\vec{H}^{(1)}\cdot d\vec{l}}. \tag{16}$$

A deep inspection of the above expression shows that there are several interesting characteristics. First, we can see that if the region (2) does not exist and the effective permittivity $\varepsilon_1$ equals 0, the reflection coefficient will vanish recovering the perfectly matched zero-index medium like discussed in the previous work [11].

Second, as one of the most intriguing features, if the region (2) is a perfect electric conductor (PEC) no matter what kind of geometric shape and how large of size it is, the incident wave will be totally blocked (reflected), i.e., a "super-reflection" phenomenon. This phenomenon is due to the fact that the electric field $E^{(1)}$ in the region (1) should be zero imposed by boundary condition at the interface of inclusion, while the magnetic field is still finite, so that $\eta_0\oint_{\partial A2}\vec{H}^{(1)}\cdot d\vec{l}\Big/E^{(1)}$ approaches infinite and the reflection coefficient equal 1. To illuminate this effect, we perform the numerical simulations. In the simulations, the electric permittivity and magnetic permeability of the region (1) medium are both characterized by Drude medium models

$$\varepsilon_1 = 1-\frac{\omega_{pe}^2}{\omega(\omega+i\omega\gamma_e)}, \qquad \mu_1 = 1-\frac{\omega_{pm}^2}{\omega(\omega+i\omega\gamma_m)}. \tag{17}$$

For simplicity, we assume that the parameters of the Drude medium models are identical, i.e. $\omega_{pe} = \omega_{pm} = \omega_p$ and $\gamma_e = \gamma_m = 1\times 10^{-5}\omega_p$. This means that the refraction index of such medium must vanish at the operating frequency $\omega = \omega_p$. We suppose that $\omega_p a/c = 2\pi$. All the results of numerical simulations in this work are computed by the commercial COMSOL Multiphysics based on the finite element method. The solid line (denoted by S1) of Fig. 2 shows the transmission coefficient of the region (1) embedded with a circular PEC barrier as function of normalized frequency (results for arbitrary shape PEC barriers are similar). As comparison, in Fig. 2, we also plot the transmission spectra when the circular PEC barrier (the region (2)) is replaced by the ZIM material (dashed line, denoted by C1), and when the ZIM in the region (1) is replaced by air (dash-dot line, denoted by C2). As expected, we see that for the C1 case the transmission coefficient spectrum shows that the wave is perfectly tunneled, whereas for the S1 case at the plasma frequency the transmission characteristic has a distinct dip indicating that the incident wave is totally blocked. Fig. 3 shows the resulting electric-field ($E_z$) and the magnetic-field ($H_x, H_y$) distribution for the S1 case at $\omega = \omega_p$. The simulated fields show that inside the region (1) the electric fields indeed vanish, while the magnetic fields are rotating around the PEC barrier. It is easy to obtain that $\eta_0 \oint_{\partial A2} \vec{H}^{(1)} \cdot d\vec{l} \big/ E^{(1)} \to \infty$, and no transmission signal will be found at $\omega = \omega_p$.This is in consistent with the above discussions. Here we just discuss the normal incidence case. Actually, such effect is still valid for arbitrary incident angle. In addition, according to the duality principle, the results can also be extended for TM mode as long as the inclusion is replaced by any PMC structure.

Another interesting feature is that if an arbitrary shaped object coated with a thin PMC shell is embedded in the ZIM, the object will be hidden by the medium, similar to cloaking devices [3, 4]. The reason is that the magnetic field at the surface of the inclusion is perpendicular to the interface $\partial A_2$ imposed by the boundary condition of PMC, at the same time, the electric field is finite constant, the value of $\eta_0 \oint_{\partial A2} \vec{H}^{(1)} \cdot d\vec{l} \big/ E^{(1)}$ must vanish and the wave can totally tunnel through the region (1) like propagating at matched zero-index medium. We know that in the matched zero-index medium the wavelength of ray is extremely large, the wave is permitted to tunnel through the ZIM due to super coupling effect, and there are no phase differences between the input and output wave fronts. For this case, a "diamond" with a thin PMC shell is chosen as the object to be hidden in our numerical simulations. The solid line of Fig. 4 shows the transmission coefficient as function of normalized frequency for the ZIM including the object case (denoted by S2). For a comparison, the transmission spectrum where the region (1) is replaced by air while the "diamond" object kept is plotted in the same figure as dash-dot line (denoted by C3). The results of Fig. 4 reveal that at $\omega = \omega_p$, the wave does, in effect, completely tunnel through the ZIM, which is just the necessary condition for cloaking device. The resulting electric field and magnetic field distribution for this case are plotted in Fig. 5. We note that inside the region (1) the magnetic field at the surface of object is always perpendicular to the interface (i.e. the boundary $\partial A_2$), and the electric field takes a constant, so that the value of $\eta_0 \oint_{\partial A2} \vec{H}^{(1)} \cdot d\vec{l} \big/ E^{(1)}$ will be strictly zero. It is as if the ZIM is empty at $\omega = \omega_p$, the illuminating wave can totally tunnel through it due to super coupling effect without phase variation. Consequently, we prove that the ZIM can be used to conceal objects, when the object is covered by a thin PMC shell. In the same manner, the ZIM cloak is also available for the TM wave when the object coating is replaced by PEC lining. Whereas, this cloaking effect is only useful for transverse electromagnetic (TEM) mode, because of the wave with any oblique incident angle cannot fully pass though ZIM material due to impedance mismatch.

In summary, we have demonstrated the phenomena of super-reflection and cloaking based on the ZIM materials. The ideas are numerically confirmed by full wave simulations. In practice, some natural materials (e.g., some noble metals) have the permittivity near zero at the infrared and optical frequencies. Meanwhile, the ZIMs can also be made by metamaterials using some specific resonance structures, which can in principle work at any frequency regime. Thus, there is practical possibility of realizing such ZIM's super-reflection and cloaking phenomena. This work is supported by the Swedish Foundation for Strategic Research (SSF) through the Future Research Leaders program and the Swedish Research Council (VR).

## References

* Corresponding author. E-mail: min@kth.se

# Figures

Fig. 1 (Color online) Geometry of a 2D waveguide structure with a ZIM material (region 1). An object (region 2) can be placed in the ZIM material.

Fig. 2 (Color online) Transmission coefficients as a function of the normalized frequency. S1: The ZIM region embedded with the circular PEC barrier; C1: the circular PEC barrier replaced by the ZIM material; C2: the ZIM in the region (1) replaced by air. The inset shows the geometry of the S1 case.

Fig. 3 (Color online) The resulting electric field ($E_z$) and magnetic fields ($H_x, H_y$) distribution for the S1 case at $\omega = \omega_p$.

Fig. 4 (Color online) Transmission coefficients as a function of the normalized frequency. S2: The ZIM region including the concealed object; C3: the ZIM in the region (1) replaced by air. The inset shows the geometry of the ZIM cloak concealing a “diamond”.

Fig. 5 (Color online) The resulting electric field ($E_z$) and magnetic fields ($H_x, H_y$) distribution for the S2 (ZIM + Object) case at $\omega = \omega_p$.

Figure 1

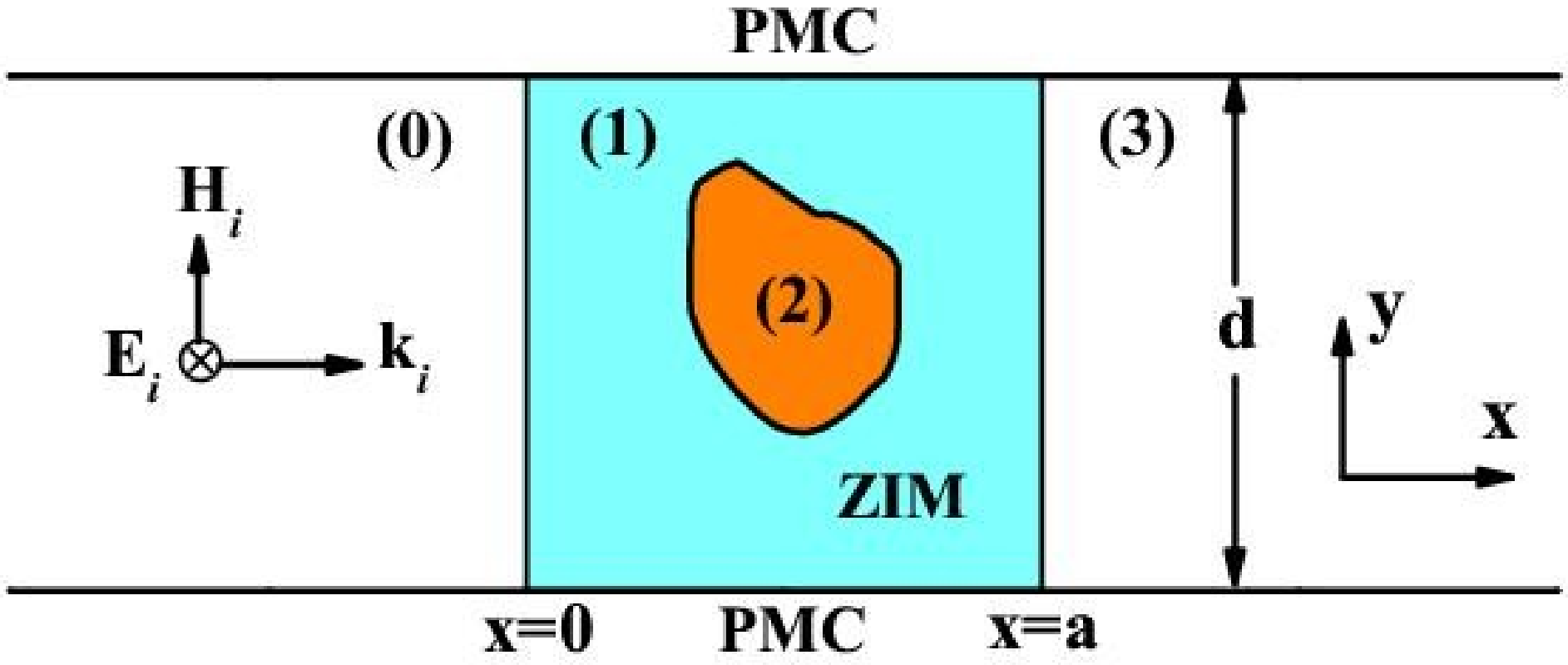

Figure 2

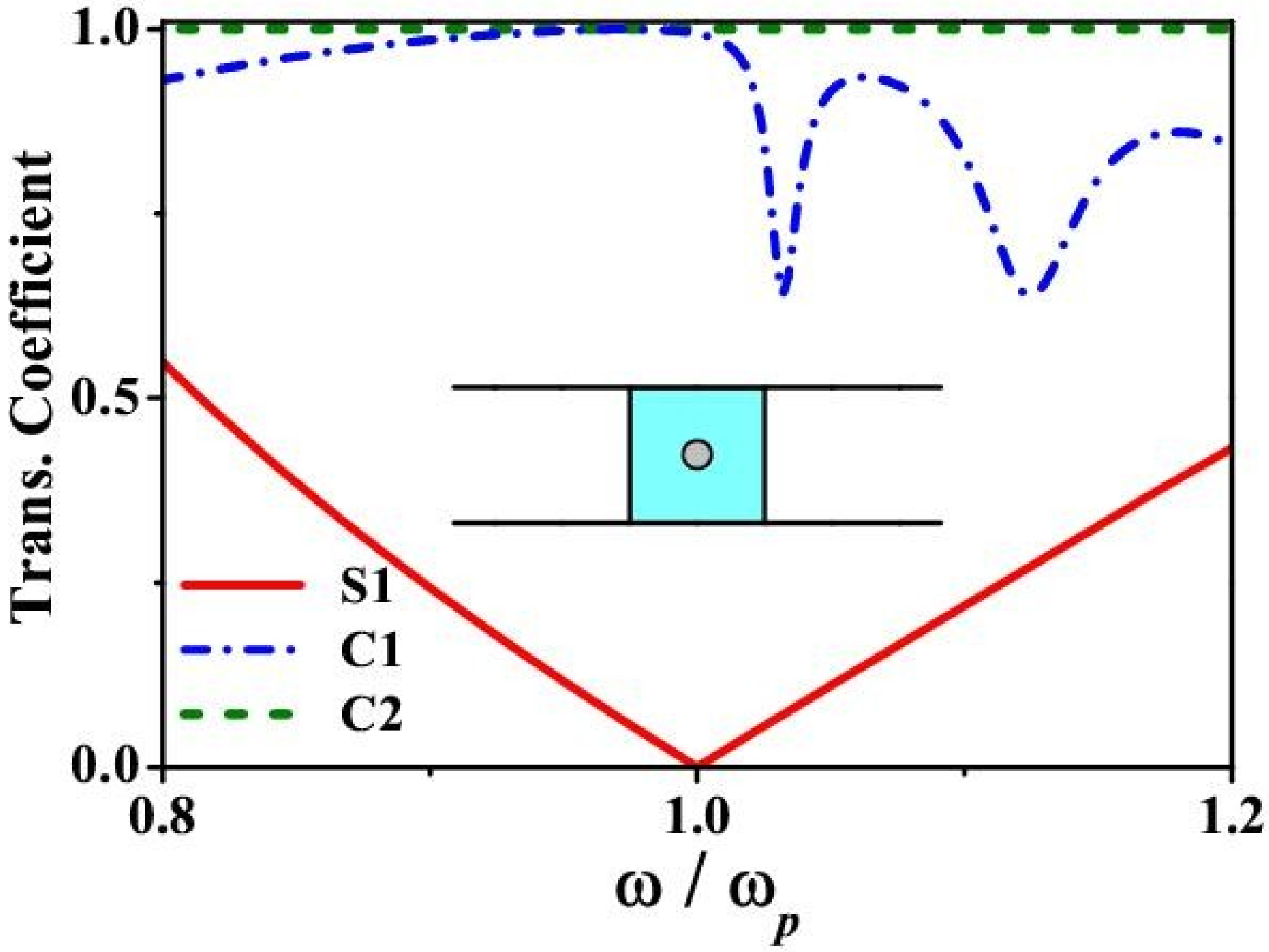

Figure 3

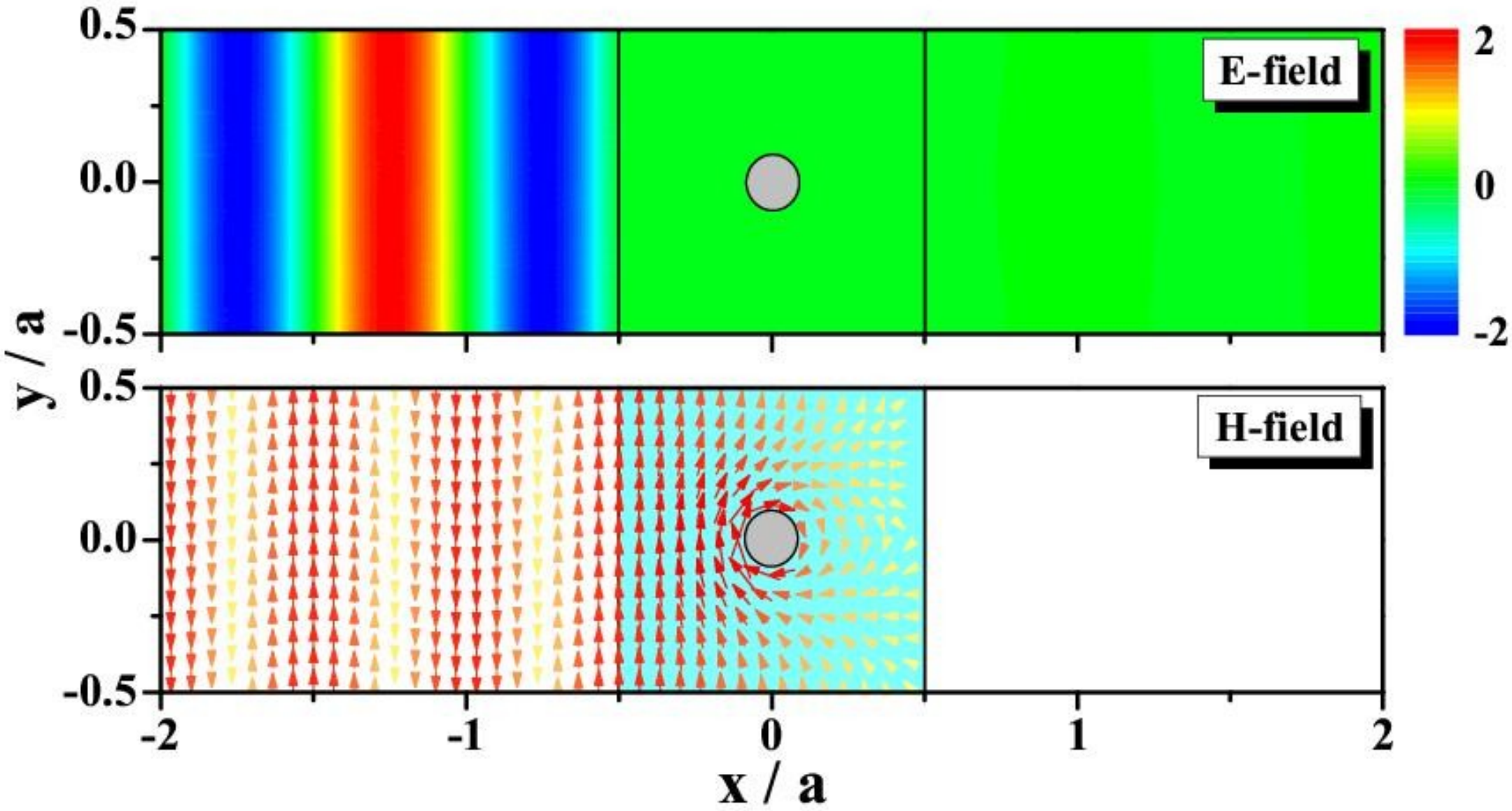

Figure 4

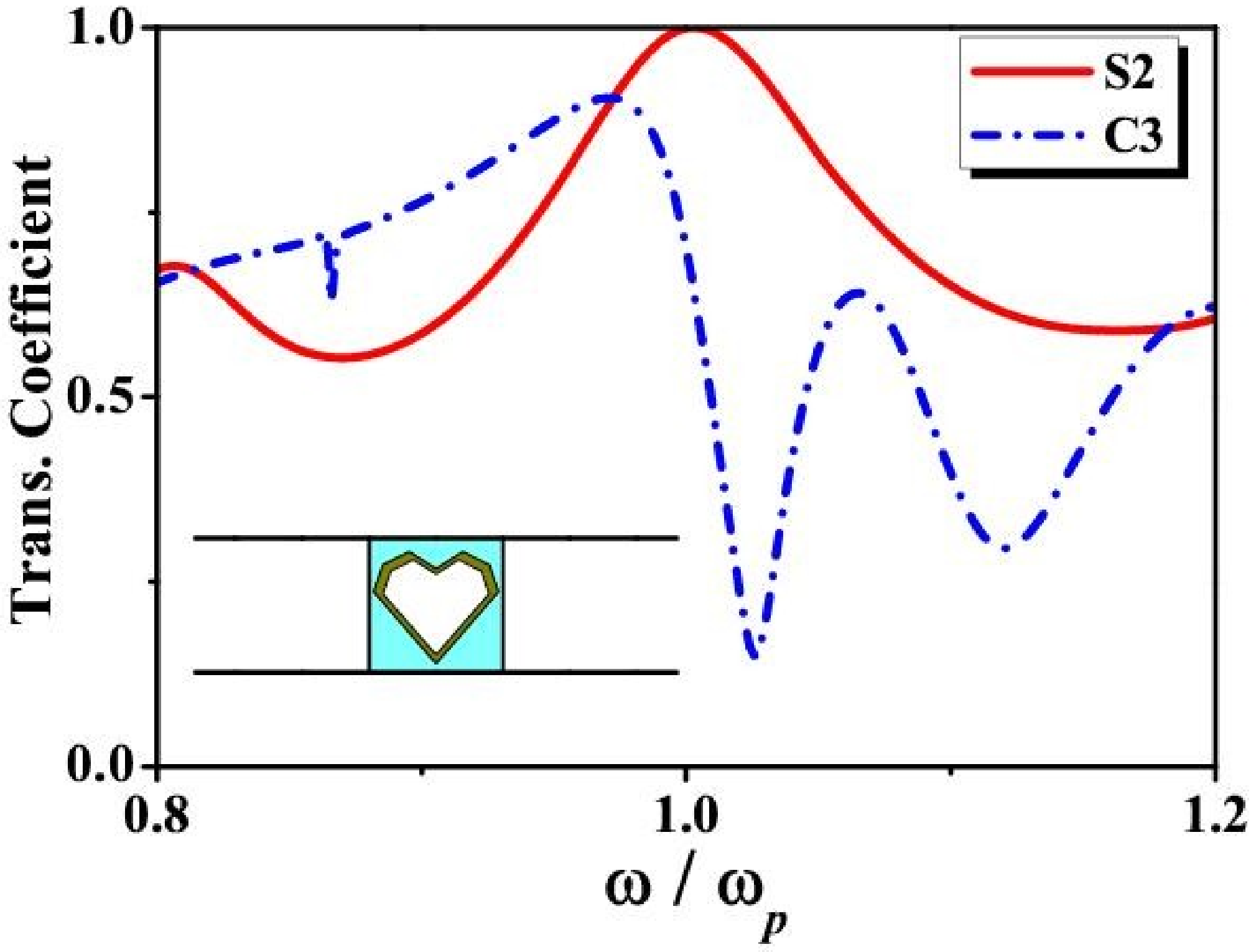

Figure 5

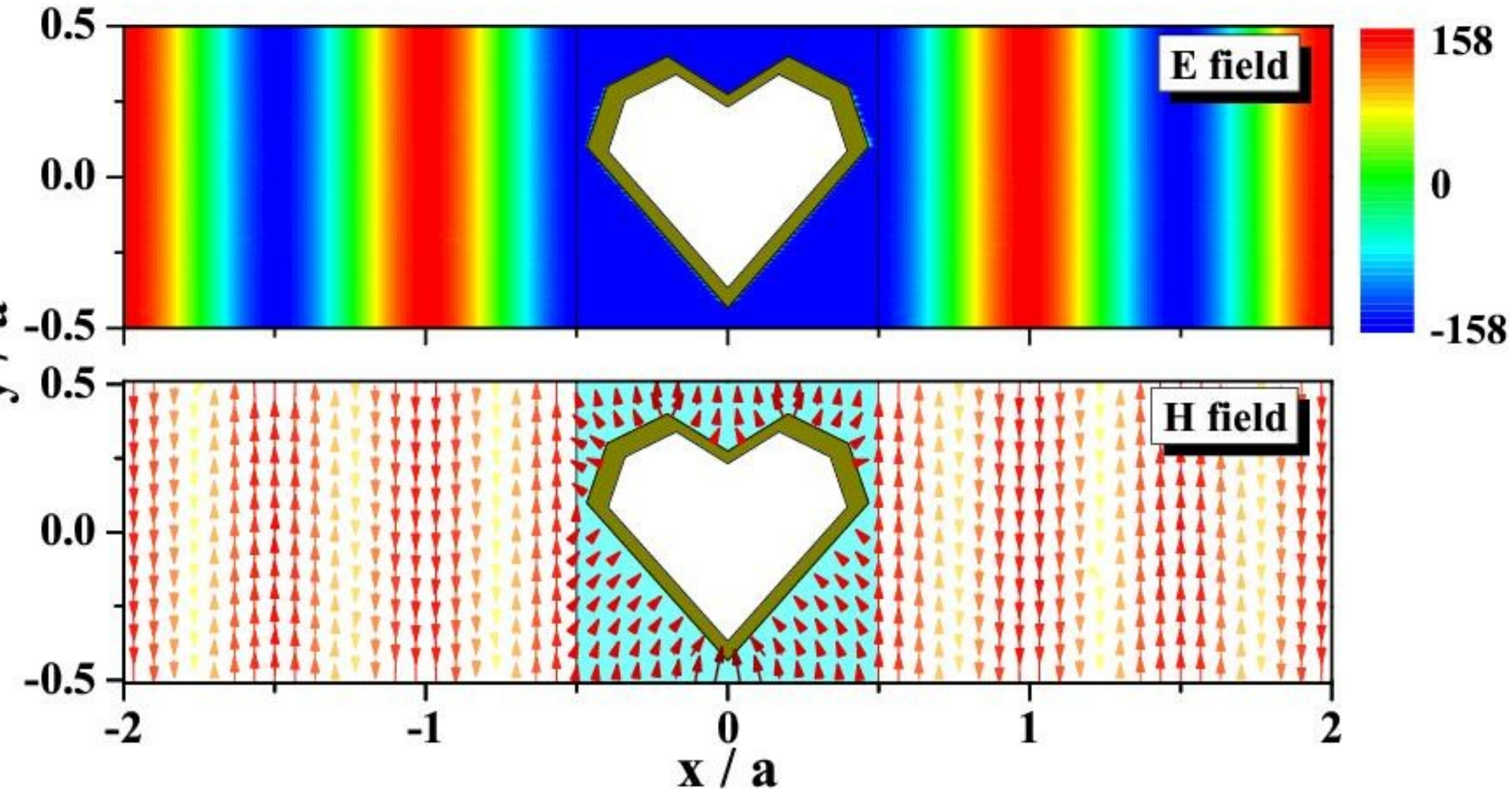